# Jumping over the paywall: Strategies and motivations for scholarly piracy and other alternatives

Francisco Segado-Boj
Complutense University of Madrid
https://orcid.org/0000-0001-7750-3755

Juan Martín-Quevedo
Universidad Rey Juan Carlos
https://orcid.org/0000-0003-1005-0469

Juan-Jose Prieto-Gutiérrez
COYSODI (Communication and Digital Society) Research Group, Universidad Internacional de la Rioja, Madrid, Spain
https://orcid.org/0000-0002-1730-8621

# Jumping over the paywall: Strategies and motivations for scholarly piracy and other alternatives

**Abstract**
Despite the advance of the Open Access (OA) movement, most scholarly production can only be accessed through a paywall. We conduct an international survey among researchers (N=3,304) to measure the willingness and motivations to use (or not use) scholarly piracy sites, and other alternatives to overcome a paywall such as paying with their own money, institutional loans, just reading the abstract, asking the corresponding author for a copy of the document, asking a colleague to get the document for them, or searching for an OA version of the paper. We also explore differences in terms of age, professional position, country income level, discipline, and commitment to OA. The results show that researchers most frequently look for OA versions of the documents. However, more than 50% of the participants have used a scholarly piracy site at least once. This is less common in high-income countries, and among older and better- established scholars. Regarding disciplines, such services were less used in Life & Health Sciences and Social Sciences. Those who have never used a pirate library highlighted ethical and legal objections or pointed out that they were not aware of the existence of such libraries.

**Introduction**

Although the scientific journal remains the cornerstone of the scholarly communication system, it is undergoing several transformations (Herman et al., 2020). Among them, the traditional business model, which requires expensive subscriptions to the journals or fees to access individual articles, has received the most criticism. Such pay-to-read, closed, or paywalled access model is perceived by some scholars as an obstacle to the advancement of science (Nicholas et al., 2019; Segado-Boj et al., 2018) and unfair and damaging for the public interest (James, 2020).

Thus, many stakeholders in scholarly publishing are pressing on the need to progress to an Open Access (OA) model, composing what could be seen as an OA Ecosystem (OAE) over the years (Jaime et al., 2021). Authors claim ethical reasons for this change of paradigm (Van Noorden, 2018), as do some editors (Segado-Boj et al., 2017). After investigating the Springer Compact Agreement (the so-called "read and publish" journals agreement) pilot 2016–2018, Marques and Stone (2020) concluded that the prevalence of OA will increase. Funding institutions—such as the European Commission and the US National Science Foundation or private funds like the Welcome Trust or the Bill and Melinda Gates Foundation—are requiring the results of their funded projects to be published in OA. In Europe, 20 institutions from Science Europe have gone as far as organizing themselves in the cOAlition S, requiring that, as of 1st January 2021, all research funded by its members be published in OA journals or archives; this initiative was known as Plan S (Bianco and Patrizii, 2020). However, the volume of articles currently covered by the funding of the signatories of Plan S is still very small and does not make a significant impact in the overall number of OA articles (Björk, 2021).

Most scientific literature is still published in the closed access model (Piwowar et al., 2018). Nearly 75% of all scholarly documents can only be accessed through a paywall (Boudry et al., 2019). This percentage varies by discipline, from 93.6% of the documents being freely available in multidisciplinary journals to 32.3% in law, arts, and humanities (Martín-Martín et al., 2018).

This prevalence of a pay-to-read model is more detrimental to researchers with lower incomes, especially those living in the Global South, as they receive little support in the form of funding or institutional libraries with current subscriptions (Canagarajah, 2002; Curry and Lillis, 2018; Demeter, 2018; Meagher, 2021).

In this context, semi-legal or completely illegal alternative strategies have emerged to make available the paywall-protected scholarly documents. One such strategy is Bronze OA, which refers to the documents that are free to read on the publisher's website, but without an identifiable license. It includes sites such as ResearchGate that make some attempts to comply with the legal requirements—usually by making authors confirm that they are allowed to share the documents—but nonetheless are prone to include unauthorized content. Bronze OA has become the most frequent form of OA (Piwowar et al., 2018).

Another strategy is the so-called Black OA (Björk, 2017) or "Robin Hood" OA (Archambault et al., 2014; Antelman, 2017), which offers huge numbers of research documents for free (i.e., without any paywall), irrespective of copyrights, embargoes, OA status, and other considerations. One of these methods is the hashtag #icanhazPDF, which works as an encounter point where scholars can ask other users to share documents that are protected behind a paywall, regardless of whether they are the authors. However, the most usual form of Black OA are services and platforms— identified as "shadow" or "pirate" libraries—that store illegal copies of scientific documents and allow users to retrieve them (Björk, 2017). Such piracy services have become a common practice in the scholarly knowledge circuit, and are widely used in both developed and developing countries (Bohannon, 2016; Bodó, 2018). An important difference between Black and Bronze OA is that the former does not make any attempt to comply or appear to comply with the legal restrictions on accessing documents.

The most popular initiative among such services is Sci-Hub (González-Solar and Fernández-Marcial, 2019). The pressure from publishers has frequently blocked Sci-Hub websites, but it has demonstrated remarkable resilience by resurfacing with a slightly different URL and continuing to grow. In 2017, Sci-Hub provided access to "nearly all scholarly literature," which translated to 85.1% of the articles published in closed access journals, with more than 56 million references (Himmelstein et al., 2018). In 2021, Sci-Hub bragged of having 88.5 million references in its database (Sci-Hub, 2021). Articles downloaded from Sci-Hub receive more citations (1.72 times more, based on data from 12 leading journals in economics, consumer research, neuroscience, and multidisciplinary research) than those downloaded from other sites (Correa et al., 2022). Some studies surveyed how and why scholars resort to Bronze or Black OA, even before these terms were coined. Gardner and Gardner (2015) studied the usage of the hashtag #icanhazpdf, which was used on Twitter for the free interchange of articles and other scholarly documents among researchers. They concluded that most users only asked for one article that was mostly published more than five years ago, which suggests that these researchers used Twitter not as their principal way of accessing documents, but more as a way of locating difficult-to-find publications that were too old to be available through the usual methods. Cenite et al. (2009) found that these practices are used as a form of bypassing market availability limitations, while also pointing out that a sense of community and ethical compromise was involved in sharing these services among the users. Further, Gardner and Gardner (2017)—based on a reduced sample of subjects, but more diverse in national and demographic characteristics than that of Cenite et al. (2009)—pointed out that the most frequent motivations for using pirate libraries are the lack of access through legal channels and the advantage of the speed of Black OA compared to the burdensome bureaucracy of institutional channels, which can take days or weeks of procedures to obtain a document. Very few subjects suggested ideological reasons, pointing to a mainly utilitarian mindset. This is also reinforced when most users responded that they "don´t care" about copyright infringements, avoiding taking an ethical stance regarding Black OA. The findings of Björk (2017) coincided with these conclusions, as he highlighted three main reasons for Sci-Hub's popularity: ease of use, the perception that downloading articles does not entail legal risks, and most scholars finding Black OA morally acceptable.

Studies have also warned of the potential negative effects of Sci-Hub. As Sci-Hub provides pirated free access to the vast majority of scientific papers, it has been deemed to eclipse the legal modalities of OA publishing such as the Green or Gold roads (Green, 2017). Moreover, the popularity of Sci-Hub reduces the paywalled access through institutional subscriptions to scientific documents. This is a move that could theoretically lead libraries to cancel their agreements with publishers (not because the libraries themselves resort to pirate sites, but because they could receive fewer requests for purchase or subscriptions as readers resort to easier and quicker, albeit illegal, ways of accessing paywalled documents), severely damaging the incomes of traditional journals and compromising their future (Dinu and Baiget, 2019; McKenzie, 2017; Marple, 2018). In another sense, even though some researchers advocate the use and support of pirate OA as a kind of civil disobedience action (James, 2020), pirate resources have also been deemed detrimental to the OA movement. According to Couto and Ferreira (2019), as Sci-Hub and other similar services provide access to paywalled literature, researchers might perceive a lesser need for supporting OA as they can read and consult the literature they need for their research purposes; moreover, the rise of Black OA could lead to even steeper, more restricted paywalls (Novo and Onishi, 2017).

Previous studies have also highlighted the generalized use of pirate libraries across
countries and disciplines (Bohannon, 2016; Behboudi, Shamsi, and de la Fuente, 2021). In some cases, Sci-Hub provides free access to more than 90% of the research papers from India (Singh et al., 2020). Some data suggest that richer regions use pirate resources more frequently (Bodó et al., 2020; Walters, 2019), and other reports state that such platforms are more intensely used in lower-middle-income countries (Till et al., 2019). Furthermore, researchers from these countries, who make use of parallel libraries, are more able to publish in international academic journals (Buehling et al., 2022). As for disciplines, former analyses state that Sci-Hub is more used to download chemistry papers (Greshake, 2017).

**Justification and novelty**

This study introduces a global international survey that has measured the habits and reasons to use (or not use) scholarly pirate resources. Some preceding studies in this direction have been based on the released usage data from platforms such as Sci-Hub (Behboudi, Shamsi, and de la Fuente, 2021; Bohannon, 2016; Greshake, 2017; Machin- Mastromatteo, Uribe-Tirado, and Romero-Ortiz, 2016; Till et al., 2019) and Library

Genesis (Bodó et al., 2020). Such an approach restricts the scope of the data to the effective download of documents on the site, but misses information about the motivation of the users and the opinions of those who do not use such services.

Other studies regarding the use of Sci-Hub have introduced survey results but their samples were restricted to medical students (Mejía et al., 2017), limited to one institution (Duić, Konjevod and Grzunov, 2017), or recruited through a convenience sample resulting in their results being biased toward heavy users of the service (Travis, 2016). Other qualitative studies have also addressed the issue of Black OA, but they have been restricted to early career researchers (Nicholas et al., 2019).

Therefore, our study is the first to conduct a random sample survey and provide information about not only those who use scholarly pirate resources but also those who do not, and the reasons behind each decision. Our design also considers other options to overcome paywalled articles and compares them to Black OA. Further, we analyze how such attitudes and habits differ according to several factors: age, professional position, country income level, discipline, and commitment to OA.

Early career researchers worldwide have shown a mostly positive attitude toward pirate libraries (Nicholas et al., 2019), but little is known about the perception and use of these resources among older, better-established colleagues. We expect older, more senior faculty staff to show a more reticent attitude toward Black OA. Therefore, we consider the role of age and seniority as two different predictors in our model, as both of them have been separately identified as influences on, for instance, the perception of the OA publishing model (Rodríguez, 2014; Zhu, 2017).

We also consider country income level as a factor, given the conflicting evidence (Bodó et al., 2020; Till et al., 2019) on how this feature influences the use of different pirate libraries. We explore how (or if) this might be a significant predictor.

We also include discipline as an independent variable in our model, as previous studies have stated that the information-seeking behavior differs according to the scholars' different disciplines.

Finally, we introduce commitment to OA publishing to explore its role in predicting habits and motivations to use Black OA sites. As previously stated, scholarly piracy might be detrimental to the OA movement (Couto and Ferreira, 2019), but is also a kind of civil disobedience action (James, 2020). Therefore, we explore if scholars more involved in OA self-archiving are more frequent users of pirate libraries or, on the contrary, refrain from using such services.

**Research objectives**

We accordingly developed the following objectives (O) and research questions (RQ): O1: To identify the strategies used by researchers to consult articles behind a paywall, and gauge the relative importance of each strategy.
O2: To quantify the reasons that lead researchers to use or not use scholarly pirate resources.
O3: To examine the relationship of the researchers' attitudes and actions with their personal (age), professional (position), socioeconomic (income level of the country of affiliation), and academic (area of knowledge) characteristics, as well as the habit of publishing in OA.

RQ1. What strategies do researchers follow to read paywalled articles?
RQ1.1. How do age, position, country income level, discipline, and commitment to OA predict what strategies are followed?

RQ2. What are the reasons behind the use of scholarly pirate resources?
RQ2.1. How do age, position, country income level, discipline, and commitment to OA predict such reasons?
RQ3. Why do scholars choose not to use scholarly pirate resources?
RQ3.1. How do age, position, country income level, discipline, and commitment to OA predict such reasons?

**Methods**

We collected the data from the Scopus bibliographic database, considering the authors of the articles published in scholarly journals across the world—not restricted to one particular country or region—as our population of interest, based on authorship rather than academic affiliation (i.e., it includes authors outside the university setting).

Given the technical difficulties in downloading the dataset of the total scholarly production of two years, we selected a stratified random strategy for selecting the survey participants. Thus, instead of randomly approaching the whole universe of researchers and authors in a given bibliometric database, we restricted our sample to randomly selected journals in four main disciplines and then addressed the corresponding authors of the journals. Thus, participants were chosen through a two- step procedure. We first selected a random group of journals from different disciplines and later retrieved the contact information for the corresponding authors of the papers published in those journals in 2019–2020. Thus, we took all the corresponding authors of published manuscripts in Scopus (2020 edition, the latest available when this study was developed) indexed journals as our universe of study.

Journals in the 2020 Scopus edition were categorized into four big groups by subject areas according to the SCImago Journal Rank. We added a fifth category to include journals from Africa and Latin America, to expand the number of responses from a non- Northern/Western perspective. For each category, a sample of journals was selected to reach a 95% confidence interval and a +/-5% margin error (see Table 1).

Subsequently, we directly downloaded from Scopus the information for the papers published in each journal in the considered time frame, including the corresponding authors' email when available. We gathered 88,892 authors' emails to which we manually sent the invitation to participate in the survey from our institutional emails via a self-administered online form that automatized the data compilation (Google Forms). From April 25 to July 10, 2021 we collected 3,304 valid responses. The response rate (4%) was higher than that of previous surveys addressed to global and massive universe of study unrestricted to a single discipline or country (see, e.g., Kień´c, 2017).

The study design, self-administered form, and mandatory informed consent form were approved by the Institutional Review Board of one of the authors' universities (International University of La Rioja – Code: PI:004/2021). Participants' responses were collected and analyzed in an aggregated manner so that they could not be individually identified.

**Table 1.** Sampling details

| | Sources (articles with individual bibliographic records in Scopus) | Sources (unique) | Sampled journals | Retrieved emails | Emails (unique) |
|---|---|---|---|---|---|
| Arts & Humanities | 4182 | 3501 | 353 | 6156 | 5955 |
| Life Sciences | 5927 | 4908 | 357 | 21673 | 18395 |
| STEM | 14766 | 10112 | 371 | 41640 | 37244 |
| Social Sciences | 11602 | 9685 | 371 | 19422 | 18800 |
| Africa & LATAM | 1199 | 1199 | 292 | 6062 | 4996 |
| | | | | Total unique emails | 82603 |

**Measurements**

Participants were required to indicate their age, gender, and current professional position. They also had to specify the country in which the institution they were affiliated with was based from a list of countries specified in the SCImago institutional rankings. This information was later recoded to the country income level information in the latest World Bank Report: low-income, lower-middle, upper-middle, and high-income countries. Due to the low number of responses from low-income countries, we aggregated the low and lower-middle categories. Further, from the categories in the SCImago institutional rankings, participants had to choose the main sector of their institution (Government/Health/Non-Profit/Private/University & Higher Education). The form included a warning for those affiliated with more than one center, who were asked to provide the information regarding their primary affiliation—the one where they developed more of their work.

Participants were also required to choose their main subject area of research from the Scopus categories. Their responses were later aggregated into four main disciplines: Life & Health Sciences (including Medicine; Biochemistry, Genetics, and Molecular Biology; Dentistry; Health Professions; Immunology and Microbiology; Neuroscience; Nursing; Pharmacology, Toxicology, and Veterinary), Science, Technology, Engineering, and Mathematics (STEM) (including Agricultural and Biological Sciences, Chemical Engineering, Chemistry, Computer Science, Decision Sciences, Earth and Planetary Sciences, Energy, Engineering, Environmental Science, Materials Science, Mathematics, Multidisciplinary, and Physics and Astronomy), Social Sciences (Business, Management, and Accounting; Economics, Econometrics, and Finance; Psychology and Social Sciences), and Arts & Humanities (Arts and Humanities).

Another set of questions gathered information about the participants' habits regarding OA publishing and making their research freely available to other researchers. First, the questionnaire asked, "How often do you upload your published manuscripts or other research documents to a repository so that they can be freely downloaded by other researchers?" (Never/Infrequently/Occasionally/Whenever the publisher rights of the journal to which I submitted the manuscript allows me to do it/Always, even though the publisher rights do not allow me to do it).

Further, the participants were given a situation, "Imagine you are interested in reading a document, but you only find a version behind a paywall, or which is not under your institutional subscription," with different options. They had to rate on a Likert-scale how often they followed each of the following proposed pathways to the paywalled documents:

I look for an OA version of the document (through Google or an academic search engine) or through services like Unpaywall.
I use pirated document repositories like Sci-Hub, Library Genesis, or 91lib.
I ask colleagues from other institutions for the paper.
I write to the corresponding author requesting a copy of the document.
I only use the data or information in the abstract and stop looking for the document.
I specifically ask my institution's library to buy a copy of the document or get it through interinstitutional loans.
I pay to access the document with my research funding or own money.
We included not only pirate libraries, but also other strategies followed by the users to acquire documents they were unable to access by institutional means (Łuczaj and Holy- Łucza, 2020).

Finally, the form included questions about the reasons scholars use (or not use) scholarly pirate resources. Following Travis (2016), participants were asked to indicate the "primary reason for using Sci-Hub or other pirated document repositories (Library Genesis, etc.)" by choosing one of the following answers: "I don't have any access to the papers," "It's easier to use than the authentication systems provided by the publishers or my libraries," or "I object to the profits publishers make off academics." Participants also could select the answer "I don't use pirated document repositories" in which case they had to specify why they did not use these resources: "I didn't know they existed," "I find it difficult (the process is confusing, I get lost in the changes of domain, or other)," "I think it is unethical and unlawful," or "I think it damages the Open Access movement."

**Data analysis**

To identify differences among groups, we applied a non-parametric analysis of variance (ANOVA) test (Kruskal–Wallis) as the compared values followed a non-normal distribution ($p<.001$ in every case in the Shapiro-Wilks Test). Dwass-Steel-Critchlow- Fligner pairwise comparisons were run to identify the significant differences found between the groups. For the sake of brevity, we only detail the comparisons where a significant difference ($p<.001$) was identified. The W statistic is calculated as the differences between the number of standard errors separating the observed sample mean and the mean predicted by the null hypothesis. The larger or smaller the W value, the more the confidence to reject the null hypothesis (Navarro, 2013).
We also designed a regression model to identify the predictor role of the considered independent variables. As the dependent variables are ordinal values, we adopted an Ordinal Logistic Regression (OLR) test.
For this OLR test and the Kruskal–Wallis ANOVA test in RQ2 and RQ3, we converted the variables of age, country income level, position, and commitment to OA into ordinal variables. For the OLR, these values were considered the predictors, but in the Kruskal–Wallis test, they were taken as the dependent variables, and the reasons to use or not use pirate resources were the grouping variables.

Finally, we ran a chi-square test of independence to look for relationships among the categorical variables regarding the researchers' reasons for using or not using pirate resources and their different disciplines.

All tests were run through the R programming language. For the OLR tests, we used the MASS package (Ripley et al., 2018). To ensure that only large effects were taken into account, we considered p values equal to or lower than .001 as significant.

Figures 1–4 represent the distribution of responses in each category as boxplots. The thick horizontal line in the middle of each box stands for the median, and the box itself varies from the 25th to the 75th percentile; that is, it includes the second and third quartile. The vertical lines below and above the box represent the lowest and highest quartiles. The circular points indicate extreme values outside the interquartile range (outliers).

As we set the significance threshold at p<.001, we do not specify when reporting statistical significance.

**Sample characteristics**

Most participants were between 36 and 50 years old, belonged to STEM disciplines, were affiliated with institutions in high-income countries, worked in the higher education sector, worked in a tenured position, and reported a high commitment to OA self-archiving (see Table 2). The numbers between brackets indicate the rank attributed to each ordinal category.

Table 2. Sociodemographic features of the sample

| Age | | Counts | % of Total |
|---|---|---|---|
| | (1) 25 or younger | 34 | 1.0 % |
| | (2) Between 26 and 35 | 687 | 20.8 % |
| | (3) Between 36 and 50 | 1438 | 43.5 % |
| | (4) 51 or older | 1145 | 34.7 % |
| Discipline | | Counts | % of Total |
| | Arts & Humanities | 539 | 16.3 % |
| | Life & Health Sciences | 958 | 29.0 % |
| | STEM | 1141 | 34.5 % |
| | Social Sciences | 666 | 20.2 % |
| Position | | Counts | % of Total |
| | (1) Predoctoral fellow or PHD Student | 449 | 13.6 % |
| | (2) Untenured | 605 | 18.3 % |
| | (3) Tenure-Track | 402 | 12.2 % |
| | (4) Tenured | 1848 | 55.9 % |
| Region | | Counts | % of Total |
| | East Asia and Pacific | 335 | 10.1% |
| | Europe and Central Asia | 1428 | 43.2 % |
| | Latin America and the Caribbean | 427 | 12.9 % |
| | Middle East and North Africa | 139 | 4.2 % |
| | North America | 635 | 19.2 % |
| | South Asia | 203 | 6.1 % |
| | Sub-Saharan Africa | 137 | 4.1 % |
| Income-level countries | | Counts | % of Total |
| | (1) Low income | 30 | 0.9 % |
| | (1) Lower-middle | 455 | 13.8 % |
| | (2) Upper-middle | 670 | 20.3 % |
| | (3) High-Income | 2149 | 65.0 % |
| Sector | | Counts | % of Total |
| | Government | 380 | 11.5 % |
| | Health | 92 | 2.8 % |

|  | Non-Profit | 95 | 2.9 % |
|---|---|---|---|
|  | Private Company | 90 | 2.7 % |
|  | University- Higher Education | 2647 | 80.1 % |
| Commitment to OA |  | Counts | % of Total |
|  | (1) Never | 385 | 11.7 % |
|  | (2) Infrequently | 322 | 9.7 % |
|  | (3) Occasionally | 619 | 18.7 % |
|  | (4) Whenever the publisher rights of the journal I submitted the manuscript allows me to do it | 1638 | 49.6 % |
|  | (5) Always, even though the publisher rights do not allow me to do it | 340 | 10.3 % |

**Results**

We divided this section into three subsections, one for each research question. In each subsection, we first discuss the descriptive results and, subsequently, introduce the outcome of the statistical tests applied in each case.

What strategies do researchers follow to read paywalled articles?
Globally, the most common pathway to overcome paywalled articles (see Figure 1) was trying to find an OA version of the document (avg=3.95, SD=1.16), followed by social alternatives such as asking colleagues from other institutions (avg=2.8, SD=1.22) or reaching out to the corresponding author for a copy of the document (avg=2.71, SD=1.21). Piracy was far less common (avg=2.5, SD=1.58). The least frequent options were interinstitutional loans (avg=2.07, SD=1.26) and paying with one's own money (avg=1.28, SD=0.64). The full disaggregated results are available as supplementary material at https://doi.org/10.6084/m9.figshare.18798998.

Figure 1. Frequency distribution of strategies for overcoming paywalls

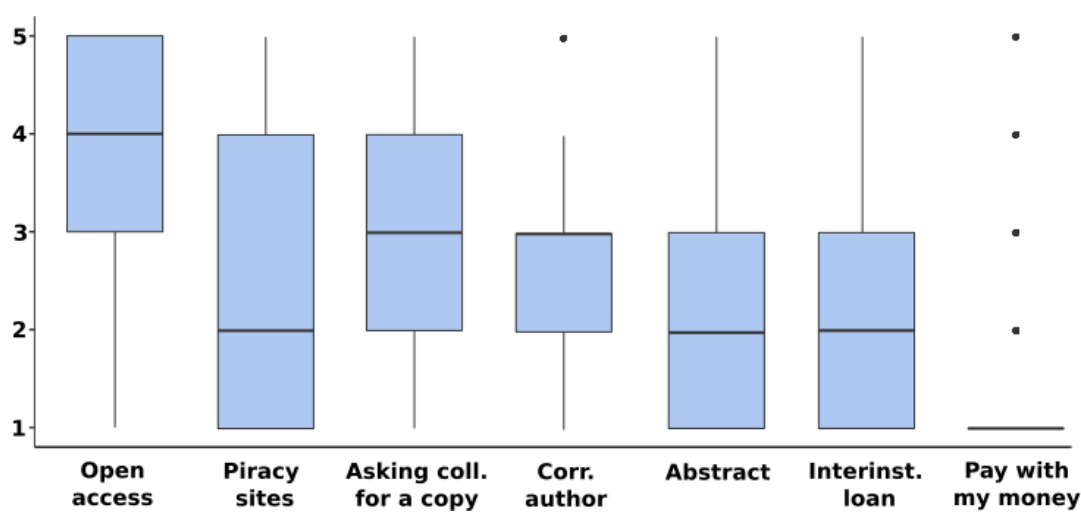

The Kruskal–Wallis test identified the effects among disciplines in the cases of OA (H(3)=34.7, p<.001), interinstitutional loan (H(3)=34.6, p<.001), and paying with own money (H(3)=16.7, p<.001). According to Dwass-Steel-Critchlow-Fligner pairwise comparisons, researchers from Arts & Humanities significantly relied more frequently on OA versions (W=5.1888, p=.001) to skip paywalls (avg=4.97) than their colleagues from Life & Health Sciences (avg=3.83). Life & Health scientists also significantly turned to OA, but less frequently (W=-6.884) than researchers from Social Sciences (avg=4.14). However, interinstitutional loans seem significantly more common (W=7.66) in Social Sciences (avg=2.30) than in Life & Health Sciences (avg=1.96).

According to the Kruskal–Wallis and post-hoc tests, paying with one's own money was more common (W=5.47) in Arts & Humanities (1.37) than in Life & Health Sciences (avg=1.22); however, the distribution in Figure 2 was identical for all four disciplines. Given the high sample sizes, it could be possible that the statistics identified differences that were practically irrelevant. We calculated the effect size ($\varepsilon^2$=0.00506) and discarded the existence of such differences given its low value.

Figure 2. Frequency distribution of strategies for overcoming paywalls by discipline

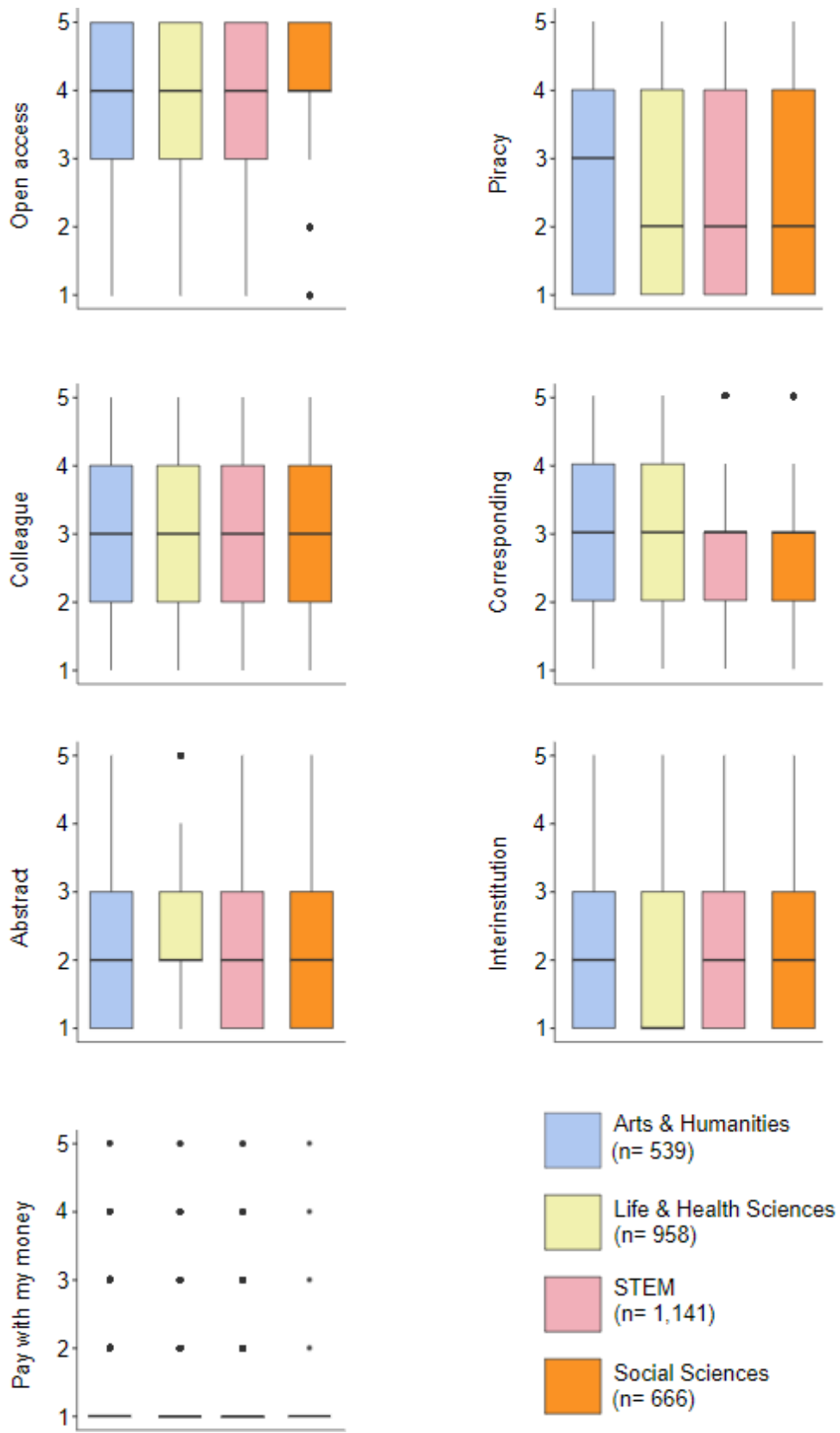

As for the use of OA, even though the median was the same in the four disciplines (see Figure 2), responses in Social Sciences were distributed in higher quartiles. Regarding interinstitutional loans, Life & Health Sciences shows the lowest median.

The OLR models were statistically significant for the different dependent variables: Open Access ($\chi^2$(4, N=3304)=124, R²McF=0.0138), pirate resources ($\chi^2$(4, N=3304)=505, R²McF=0.0525), asking a colleague ($\chi^2$4, N=3304)=52.7, R²McF=0.00514), asking the corresponding author ($\chi^2$(4, N=3304)=81.7, R²McF=0.00805), reading the abstract ($\chi^2$(4, N=3304)=33.7, R²McF=0.0035), interinstitutional loan ($\chi^2$(4, N=3304)=133, R²McF=0.0147), and pay with own money ($\chi^2$(4, N=3304)=77.9, R²McF=0.0180).

As expected, commitment to OA positively predicted the search for OA versions. According to the OLR (Table 3), participants who reported that they self-archived their articles and other results in OA repositories also searched more frequently for OA to jump paywalled documents. Age was identified as negatively related to OA, meaning that the older the researcher, the less probable it was they would look for OA documents. OA commitment also stood as a positive predictor of the use of pirate resources. The more a researcher followed self-archiving practices, the more they used pirate libraries. Younger scholars and those in low-income countries turned more frequently to piracy websites. Asking a colleague for a copy of the paper was also negatively predicted by country income level, this pathway being more common in low- income countries. This option was predicted positively by age—being more frequent among older participants—and commitment to OA. Reading just the abstract was an alternative negatively predicted by country income level (researchers from richer countries followed this habit less frequently) and commitment to OA: Those with less intense self-archiving practices followed this path more frequently.

Interinstitutional loan was predicted by country income level (being more common in richer countries) and age (being more common among older scientists). Commitment to OA played a negative role, as it seemed to deter the participants from this option.

Finally, older researchers were more prone to pay with their own money for accessing the paper. Scientists from low-income countries also seemed to choose this option more frequently than their colleagues from richer countries.

Table 3. OLR coefficients for the different strategies to overcome paywalls

| Dependent Variable | Predictor | Estimate | SE | Z | p |
|---|---|---|---|---|---|
| Open Access | Income | 0.0216 | 0.0437 | 0.494 | 0.621 |
| | Position | -0.0416 | 0.0334 | -1.247 | 0.212 |
| | Age | -0.2364 | 0.0495 | -4.778 | <.001 |
| | Commitment to OA | 0.2450 | 0.0281 | 8.734 | <.001 |
| Shadow resources | Income | -0.4028 | 0.0430 | -9.37 | <.001 |
| | Position | -0.0953 | 0.0329 | -2.89 | 0.004 |
| | Age | -0.6831 | 0.0503 | -13.58 | <.001 |
| | Commitment to OA | 0.2313 | 0.0293 | 7.89 | <.001 |
| Asking a colleague | Income | -0.30655 | 0.0428 | -71545 | <.001 |
| | Position | -0.00424 | 0.0321 | -0.1321 | 0.895 |
| | Age | -0.01226 | 0.0477 | -0.2569 | 0.797 |
| | Commitment to OA | -0.00166 | 0.0272 | -0.0610 | 0.951 |
| Asking the corresponding author | Income | -0.22215 | 0.0426 | -5.218 | <.001 |
| | Position | 0.00894 | 0.0322 | 0.278 | 0.781 |
| | Age | 0.28017 | 0.0481 | 5.821 | <.001 |
| | Commitment to OA | 0.11518 | 0.0275 | 4.183 | <.001 |
| Reading the abstract | Income | -0.1578 | 0.0431 | -3.662 | <.001 |
| | Position | -0.0439 | 0.0326 | -1.347 | 0.178 |
| | Age | 0.0422 | 0.0480 | 0.879 | 0.379 |
| | Commitment to OA | -0.1180 | 0.0272 | -4.329 | <.001 |
| Interinstitutional loan | Income | 0.3596 | 0.0459 | 7.83 | <.001 |
| | Position | 0.0553 | 0.0340 | 1.63 | 0.104 |
| | Age | 0.2393 | 0.0504 | 4.75 | <.001 |
| | Commitment to OA | -0.0943 | 0.0283 | -3.34 | <.001 |
| Pay with my money | Income | -0.28339 | 0.0591 | -4.797 | <.001 |
| | Position | -0.10340 | 0.0464 | -2.229 | 0.026 |
| | Age | 0.54194 | 0.0716 | 7.572 | <.001 |
| | Commitment to OA | 0.00828 | 0.0395 | 0.209 | 0.834 |

What are the reasons behind the use of scholarly pirate resources?

More than half the participants stated that they used pirated document repositories (see Table 4), although for different motives, most frequently because of not having access to the documents. Other motives, such as being easier to use than getting legal access or a politically motivated stand against publishers, were far less common. The full disaggregated results by categories are available at https://doi.org/10.6084/m9.figshare.18800555.v1.

Table 4. Frequency of the reasons for using pirated document repositories

| | n | % |
|---|---|---|
| I don't use pirated document repositories | 1430 | 43.3 |
| I don't have access to the papers | 1188 | 36 |
| It's easier to use than the authentication systems provided by the publishers or my libraries | 338 | 10.2 |
| I object to the profits publishers make off academics | 238 | 7.2 |
| Other | 110 | 3.3 |

A chi-square test of independence found a moderate association that did not reach our significance threshold between discipline and the reasons to use pirate resources—X2 (12, N = 3304) = 25.7, p = .012. The comparison between the expected and observed counts for this test is available at https://doi.org/10.6084/m9.figshare.19009109.v1.

The Kruskal–Wallis ANOVA found significant differences for each group among the levels of age (H(4)=279.1), position (H(4)=101.9), country income level (H(4)=162.7), and commitment to OA (H(4)=62.7).

The Dwass-Steel-Critchlow-Fligner (Table 5) pairwise comparison test also identified differences among the groups. Participants who did not use pirate resources were mostly affiliated with institutions in richer countries (avg=3.36) than those who reported that they used these services because they could not access papers (avg=2.37), as a protest against the profits made by the publishers (avg=2.35), or because they were easier to use (avg=2.32). The same pattern appears in the comparisons regarding age and professional position. Those who did not use pirate libraries were significantly older (avg=3.36) than those who reported reasons such as lack of access to the papers (avg=2.92), objections to the business model (avg=2.92), and pirate libraries being easier to use (avg=2.90). They were also at higher professional ranks (avg=3.33) than those who said that they used piracy websites because they could not access the papers (avg=2.92), because they were easier to use (avg=2.9), or as a protest against publishers (avg=2.96).

Table 5. Dwass-Steel-Critchlow-Fligner pairwise comparisons of country income level, age, position, and commitment to OA, according to reasons to use pirated document repositories

|  |  |  | W | p | Mean difference |
|---|---|---|---|---|---|
| Income | I don't have access to the papers | I don't use pirated document repositories | 15.780 | <.001 | -0,3 |
|  | I don't use pirated document repositories | I object to the profits publishers make off academics | -9.947 | <.001 | 0,32 |
|  | I don't use pirated document repositories | It's easier to use than the authentication systems provided by the publishers or my libraries | -12.580 | <.001 | 0,35 |
|  | It's easier to use than the authentication systems provided by the publishers or my libraries | Other | 5.506 | <.001 | -0,33 |
| Age | I don't have access to the papers | I don't use pirated document repositories | 21.421 | <.001 | -0,44 |
|  | I don't use pirated document repositories | I object to the profits publishers make off academics | -11.902 | <.001 | 0,44 |
|  | I don't use pirated document repositories | It's easier to use than the authentication systems provided by the publishers or my libraries | -14.264 | <.001 | 0,46 |
| Position | I don't have access to the papers | I don't use pirated document repositories | 133.368 | <.001 | -0,41 |
|  | I don't use pirated document repositories | I object to the profits publishers make off academics | -71.254 | <.001 | 0,37 |
|  | I don't use pirated document repositories | It's easier to use than the authentication systems provided by the publishers or my libraries | -82.660 | <.001 | 0,43 |
| Commitment to OA | I don't have access to the papers | I don't use pirated document repositories | -85.467 | <.001 | 0,27 |
|  | I don't use pirated document repositories | I object to the profits publishers make off academics | 79.069 | <.001 | -0,46 |
|  | I don't use pirated document repositories | It's easier to use than the authentication systems provided by the publishers or my libraries | 68.000 | <.001 | -0,33 |

Figure 3. Frequency distribution of country income level, age, position, and commitment to OA by reasons for using scholarly piracy sites

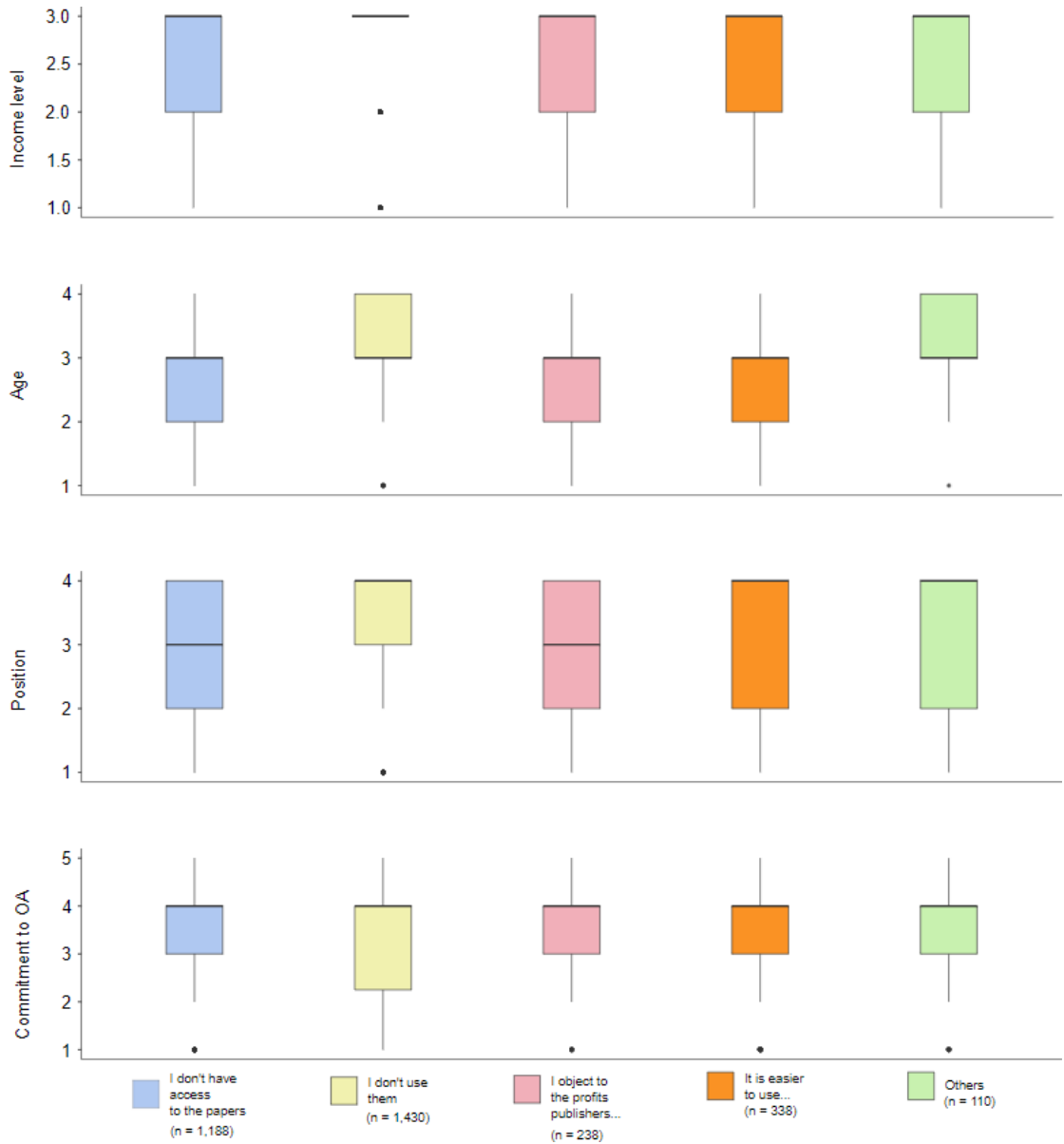

However, commitment to OA shows a different dynamic, as those who reported not using Black OA sites ranked lower in this category (avg=3.20) than those stating any other reasons such as lack of access to documents (avg=3.47), protest against publishers (avg=3.66), or ease of use (avg=3.53). This means that the users of Black OA repositories are significantly more common in poorer countries, seem to be younger, and are involved deeply in OA self-archiving. The full results for the post-hoc test are available at https://doi.org/10.6084/m9.figshare.19027400.v1.

The participants who reported that they did not use pirate resources were from high-income countries, placed in higher and more consolidated positions, and older (see

Figure 3). As for commitment to OA, the third quartile reaches lower than in the other categories.

Why do scholars choose not to use scholarly pirate resources?

The most common reason for not using scholarly pirate resources were legal and ethical concerns (Table 6), followed by ignorance of their existence. The difficulty of the process and potential damages to the OA movement have been far less frequently stated. The full disaggregated results compared by categories are available at https://doi.org/10.6084/m9.figshare.19009109.v1.

**Table 6.** Frequency of reasons for not using pirated document repositories

|  | n | % |
|---|---|---|
| I think it is unethical and unlawful | 656 | 45.9 |
| I didn't know they existed | 521 | 36.4 |
| I think it damages the OA movement | 136 | 9.5 |
| I find it difficult | 117 | 8.2 |

According to the chi-squared test (Table 7), a relationship exists between the discipline and the reasons not to use pirate resources ($\chi^2$ (9, N = 1430) = 28.3). Ethical objections were more common among STEM researchers than theoretically expected. Ignorance was also more common than expected in the cases of Life & Health scientists and their colleagues in Social Sciences. These findings imply that STEM scientists are more aware than the rest of the scholars of the moral and legal implications of pirate libraries and that such resources are less known in the Social Science disciplines and Life & Health Sciences.

Table 7. Observed and expected counts for the reasons for not using pirated document repositories and disciplines

**Table 7.** Observed and expected counts for reasons to not using pirated document repositories and disciplines

| Reasons not to use piracy | | Arts & Humanities | Life & Health Sciences | STEM | Social Sciences | Total |
|---|---|---|---|---|---|---|
| I didn't know they existed | Observed | 76 | 169 | 146 | 130 | 521 |
| | Expected | 72.9 | 159.9 | 175.6 | 112.6 | 521 |

|  | CATEGORY | | | | |
| --- | --- | --- | --- | --- | --- |
| Reasons not to use piracy | | Arts & Humanities | Life & Health Sciences | STEM | Social Sciences | Total |
| I find it difficult (the process is confusing, I get lost in the changes of domain or other) | Observed | 15 | 42 | 29 | 31 | 117 |
| | Expected | 16.4 | 35.9 | 39.4 | 25.3 | 117 |
| I think it damages the Open Access movement | Observed | 20 | 44 | 42 | 30 | 136 |
| | Expected | 19.0 | 41.8 | 45.8 | 29.4 | 136 |
| I think it is unethical and unlawful | Observed | 89 | 184 | 265 | 118 | 656 |
| | Expected | 91.7 | 201.4 | 221.1 | 141.8 | 656 |
| Total | Observed | 200 | 439 | 482 | 309 | 1430 |
| | Expected | 200.0 | 439.0 | 482.0 | 309.0 | 1430 |

The Kruskal–Wallis test identified significant differences according to income (H(3)=22.53) and age (H(3)=34.84) among those who reported different reasons for not using pirate libraries. According to Dwass-Steel-Critchlow-Fligner pairwise comparisons, participants who stated that they were not aware of the existence of such services were more frequently (W=6.513) based at institutions in low-income countries (avg=2.77) than those who stated ethical and legal objections (avg=2.59). Ethical and legal objections were also significantly more common (W=7.588) in older researchers (avg=3.46) than ignorance (avg=3.25) (Figure 4). The full results for the post-hoc test are available at https://doi.org/10.6084/m9.figshare.19027445.v1.

Figure 4. Frequency distribution of country income level, age, position, and commitment to OA by reasons for not using scholarly piracy sites

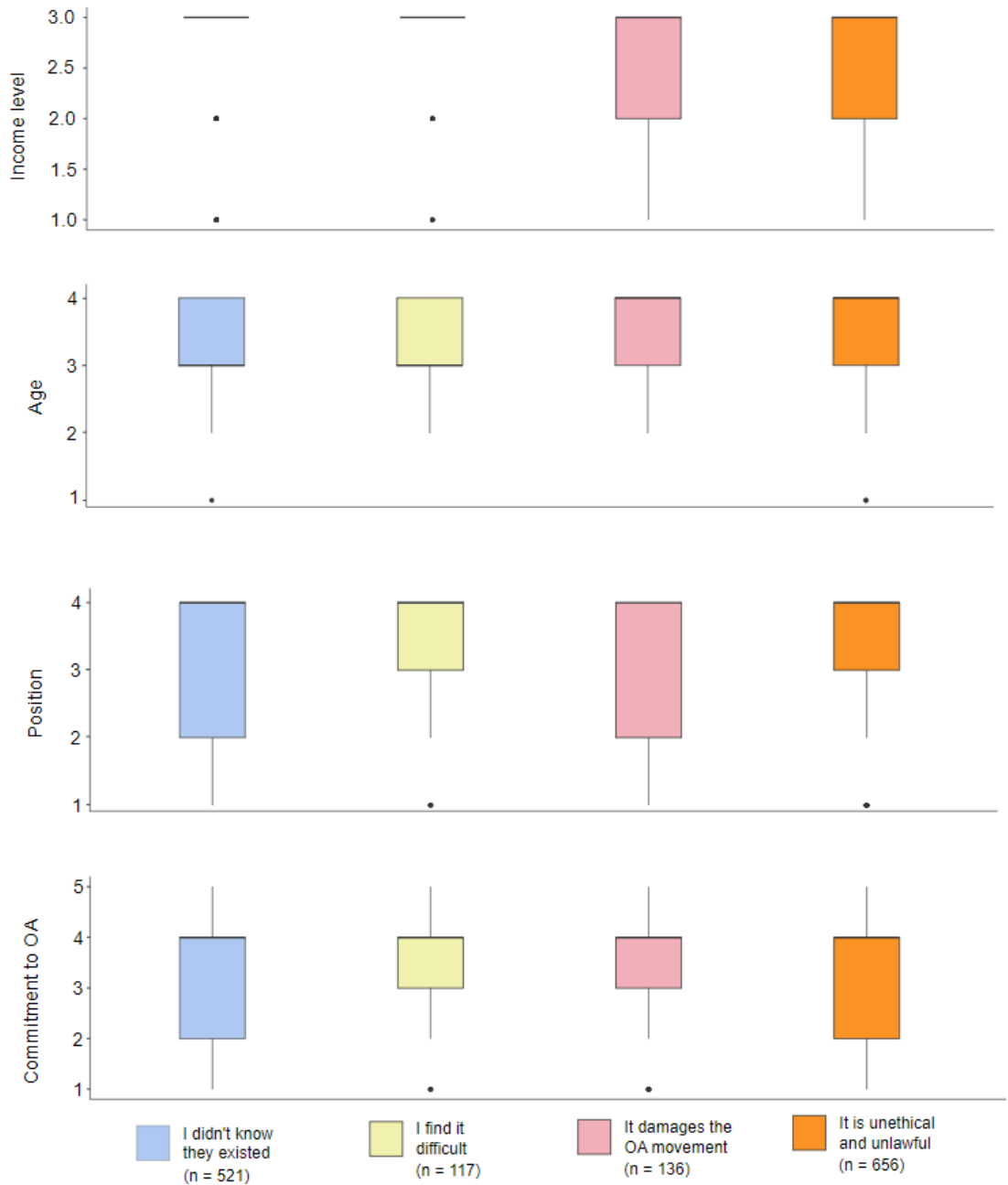

The median age level (=3) is lower for ignorance and difficulty than for that for ethical reasons or potential damage to the OA movement (=4) (see Figure 4). There are no differences in the median in the rest of the comparisons, but responses for ignorance and difficulty are more concentrated in the highest-ranked countries according to income level, with lower dispersion in the central quartiles.

**Discussion and conclusions**

Aligned with previous research (Bohannon, 2016; Behboudi et al., 2021), our results point to the extended and generalized use of pirate libraries. Further, we provide more data to understand this phenomenon better. Scholarly pirate resources are not the preferred way to access paywalled documents. Researchers repeatedly rely on Gold, Green, or Bronze OA or try to get a copy from the corresponding author or other colleagues. Piracy is a marginal pathway in comparison to those options.

We hypothesized that researchers find it easy and convenient to search for OA versions of documents such as preprints or postprints in disciplinary or institutional repositories, given that OA (via Green, Gold, or Bronze road) is a growing habit (see, e.g., Archambault et al., 2014; Piwowar et al., 2018) in both the Western developed countries (Hobert et al., 2021) and Global South (Okeji et al., 2019; Nazim and Zia, 2019), and also among different disciplines (Nazim and Zia, 2020, Hua et al., 2017). We also hypothesized that finding such OA documents through pirate resources is neither perceived as a risky process nor is it cumbersome enough to be discouraging, not only through specific software solutions like Unpaywall, but also through commonly used search engines like Google or academic specific engines like Google Scholar. Google Scholar indexes a wide array of these OA documents (Martín-Martín et al., 2018) and is an easy-to-use tool (Wu and Chen, 2014).

We also found that interinstitutional exchanges occur more commonly through informal networks (asking colleagues from other institutions for a paywalled document) than formal gateways (interinstitutional loans by libraries). Interinstitutional loans remain one of the least used alternatives to reach paywalled articles. The least frequent solution to access paywalled literature is using one's own money to pay for the required document.

We have also found that pathways for paywalled documents are more common in traditionally less-funded disciplines such as Arts and Humanities and Social Sciences than in disciplines where research grants and budgets are more relevant, including Life & Health Sciences.

Commitment to OA positively predicted turning to OA alternatives of a paywalled document. Further, researchers who frequently self-archive their papers and other results in repositories are more familiar with the OA environment, and are keener to look for OA archived preprints and postprints. Age also influences the use of OA alternatives, with younger scholars more frequently following this pathway. This is in line with the popularity of OA among early career researchers both as a publishing venue (Nicholas et al., 2020) and as a source of information discovery (Nicholas et al., 2017). The use of pirate resources was also predicted by commitment to OA. The attitude we ranked as having the highest involvement in OA self-archiving involved disregarding the terms and conditions of the publisher that originally published the document. This indifference to legal norms may influence the use of illegal resources such as pirate libraries. In practice, commitment to the OA movement does not deter the use of such pirate references, even with the potentially detrimental consequences of Black OA on the other legal roads to OA. Contrary to what Couto and Ferreira (2019) predicted, higher use of pirate libraries does not imply lower support to OA, at least in the self- archiving practices of the surveyed scholars.

In line with Nicholas et al. (2017), we found that younger researchers are keener to use piracy services. This might imply that younger scholars are incorporating pirate libraries as a natural element of their environment and information retrieval tools. However, their older colleagues are more reluctant to use such services. This finding might seriously impact the attitudes and landscape of scholarly information in the future. Our data also showed that academics from low-income countries employ scholarly piracy services more frequently. This finding allows us to generalize this trend, which has already been analyzed in some disciplines (Sagemüller et al., 2021). We understand that academics from universities and institutions with lower economic resources are keener to solve the lack of funding and gaps in their libraries by turning to Sci-Hub and other Black OA alternatives.

In this sense, our data support the findings of Cenite et al. (2009) that the lack of access to documents stands as the most relevant reason behind the use of pirate resources. Other factors, such as the ease of use or the moral acceptance of Black OA (Björk, 2017), are less frequently mentioned, supporting the hypothesis of the prevalence of a mainly utilitarian approach to Black OA. In other words, using pirate libraries is mostly understood as a solution to the problem of not having access to a particular piece of work. Although it was also hypothesized that pirate libraries might be easier to use than the authentication systems, this problem is much less frequent than the lack of access to papers. Lastly, the use of OA as a political stand or protest against the publishing industry business model is only marginal.

We have also provided evidence that scholars who did not use Black OA services were significantly older, worked in better-established positions, and were affiliated with institutions in high-income countries. This could mean that those at the core of academics mostly ignore pirate libraries. In turn, at this moment, the use of pirate libraries seems to be something more particular to the periphery, to those in less- privileged positions. Ignoring pirate libraries is a luxury that those with fewer resources at hand cannot take for granted. However, given that younger researchers are keener to use Black OA services, the situation could change in the immediate future, as early career researchers advance in their professional track and move into the center of their communities.

Regarding not using scholarly pirate resources, two reasons dominate among the participants: ethical and legal objections and lack of knowledge of the existence of such sites. Ethical objections are significantly more frequent in STEM researchers and older scholars. In contrast, ignorance is more common in low-income countries.

Thus, there are two different universes of scholars who do not use pirate libraries. The first is of researchers from rich countries, in better-established professional positions, who do not need to employ channels they consider illegal. The second includes scholars from low- and middle-income countries who are not aware of such platforms and services.

This seems like a triple burden for scholars in the periphery who already lack resources and have a constrained research budget. In addition to not being able to access knowledge through traditional or legal pathways, they also seem to be unaware of the (illegal) solutions to their problems, which add up to the difficulties in paying the article processing charge associated with many OA, well-indexed, journals (Segado-Boj et al., 2022).

In this sense, the democratizing role that Sci-Hub and other pirated document repositories claim to play has a limited effect and does not fully reach the Global South. Despite the evidence that Black OA is quite popular in low-income countries, a huge number of scholars in these nations still lack the knowledge to access Black OA. As others have already pointed out, lacking access to Sci-Hub and other pirate libraries might be detrimental to researchers in the Global South (Singh et al., 2021).

**Practical implications**

From a librarians' perspective, it is worrisome that interinstitutional loans are used very rarely. Libraries should publicize this service among their users and improve the speed of the process and procedures to request a document.

Given the relatively low rate of knowledge regarding Black OA in low-income countries, institutions and libraries could offer courses to their users on pirate libraries and their legal and moral implications.

The strategy of Sci-Hub and other pirate resources to avoid legal retaliation by opening other websites with slightly different URLs seems effective, as few scholars are discouraged from using the platform just because finding a working access URL is inconvenient or troublesome.

**Limitations and future research**

Our sample might present a bias toward Latin America and Africa. By including specific journals from these regions, we wanted to guarantee that our sample collected the perspectives and attitudes present there, which are often overlooked in the existing literature. Our goal was not to reflect the proportion of contributions to scholarly literature but to analyze the perceptions of the international community of academics. Hence, our sample might not accurately reflect the scientific output by country and regions, but instead, provide accurate data to reflect the perceptions in the Global South. In addition, we believe that including the disaggregated analysis by country income level minimizes this potential bias.

Further, as Scopus's subject coverage varies by discipline, our population might underrepresent Arts & Humanities and Social Sciences (Mongeon and Paul-Hus, 2016) due to the emphasis on journals rather than books and the underrepresentation of humanities journals within Scopus. Moreover, only including the names of the corresponding authors might result in a bias against less-experienced authors.

As for the form itself, it aggregated different ways of looking for OA versions of paywalled documents, such as search engines or other services like Unpaywall, under one option. In this regard, given the evidence that this strategy (searching for OA preprints or postprints) is quite frequent, future studies could analyze and differentiate how users look for such OA versions of paywalled documents.

Furthermore, the provided options for the professional position (tenured/untenured/tenure track) might not have been clear enough to participants unfamiliar with the Anglo-Saxon University system and those working outside higher education institutions. Moreover, we did not differentiate between "legal" and "moral" objections to the use of pirate resources.

Regarding the analysis, we are not aware of any population-characteristic data available for Scopus authors. Therefore, our tests rely on the assumption that our sample is representative of the population. Further, as we used the OLR analysis, any non-linear relationship between our predictors and outcomes might not be detected.

Finally, our survey took place in the specific context of the COVID-19 pandemic, where most of the academic publishers chose to provide free access to documents on this topic. This might have had some influence on the responses from the Life & Health Science scholars.

**Acknowledgments**

We would like to thank the researchers who participated in our survey. Without their collaboration, this work would have not been possible. We would also like to thank the anonymous reviewers for their comments and suggestions.